\documentclass[]{aa}

\newcommand{\be}{\begin{displaymath}}
\newcommand{\bn}{\begin{equation}}

\newcommand{\en}{\end{equation}}
\newcommand{\ee}{\end{displaymath}}

\newcommand{\vek}{\bf}

\newcommand{\bfm}[1]{\mbox{\boldmath$#1$}}

\authorrunning{Dendy {\it et al.}}
\titlerunning{Self-Organised Criticality and Accretion}
\begin{document}
%
\title{On the role of self-organised criticality in accretion systems}

\author{R O Dendy\inst{1}, P Helander \inst{1} \and M Tagger \inst{2}}

\offprints{R O Dendy}

\institute{UKAEA Fusion, Culham Science Centre, Abingdon, Oxfordshire, OX14
3DB, UK  \and DAPNIA/Service d'Astrophysique (CNRS URA 2052), CEA Saclay,
91191 Gif sur Yvette, France}

\date{Received ; accepted }

\thesaurus{02.01.02 - 02.13.2 - 08.02.3 - 10.03.1 - 13.25.1 - 13.25.2}

\maketitle
\begin{abstract}

Self-organised criticality (SOC) has been suggested as a potentially powerful
unifying    paradigm for interpreting the structure of, and signals from,
accretion systems.  After reviewing the most promising sites where SOC might be
observable, we consider the theoretical arguments for supposing that SOC can
occur in accretion discs.  Perhaps the most rigorous evidence is provided by
numerical modelling of energy dissipation due to magnetohydrodynamic turbulence
in accretion discs by G Geertsema \& A Achterberg ({\em A\&A}
{\bf 255}, 427 (1992)); we investigate how ``sandpile"-type dynamics arise
in this
model.  It is concluded that the potential sites for SOC in accretion
systems are
numerous and observationally accessible, and that theoretical support for the
possible occurrence of SOC can be derived from first principles.

\keywords{accretion; accretion discs -- stars: cataclysmic variables --
galaxies:
Seyfert -- X-rays: stars
               }
\end{abstract}

%

\section{Introduction}

The question whether accretion discs can be in a state of self-organised
criticality (SOC) (Bak et al. \cite{Bak:Tang}, Kadanoff et al.
\cite{Kadanoff:Nagel}) was raised explicitly by
Mineshige et al. (\cite{Mineshige:Takeuchi}), and strongly implicitly
by Bak et al. (\cite{Bak:Tang}).  Young \& Scargle
(\cite{Young:Scargle}) have raised the related question of transient chaos in
accretion systems.  A distinctive feature of SOC is flickering energy transport
with no characteristic lengthscale or time separation, displaying a $1/f$
power
spectrum; in this context, $1/f$ is
shorthand for inverse power law frequency dependence with unspecified
index.  In
SOC systems, for which mathematical sandpiles (Bak et al. \cite{Bak:Tang},
Kadanoff et al. \cite{Kadanoff:Nagel}) provide a paradigm, global transport
occurs as a result of self-organised avalanches which are triggered locally
when
the accretion of sand leads to a critical gradient being exceeded at a given
point.  This causes local redistribution of sand, which may lead to the
critical gradient being
exceeded at neighbouring points, resulting in further redistribution and,
cumulatively, to a global avalanche.  Following the avalanche, the system
returns
to a subcritical configuration; accretion then continues until it again
creates a
local excess gradient, triggering a further avalanche.  In the SOC-sandpile
paradigm, it is helpful to note that the word ``critical" is being used in two
senses: the sandpile has a critical gradient, redistribution being triggered
wherever this gradient is exceeded; and the self-organised global avalanches
which emerge from the integrated effects of local redistribution may be
scale-free, in which case they are linked to the generic field of critical
phenomena.

In mathematical models
and also some experimental realisations (Nagel \cite{Nagel:1992}, Feder
\cite{Feder:1995}, Frette et al. \cite{Frette:Christensen},
Christensen et al. \cite{Christensen:Corral}), avalanche statistics display
scale-free $1/f$ characteristics.  SOC sandpile algorithms are extremely
simple and possess the
attraction of any successful reduced system: it becomes unnecessary to
attempt to
model the detailed, and perhaps insuperably complex, microphysics of transport
in the real system.  Such an approach has been applied to the statistics of
solar X-ray
bursts (Lu \& Hamilton \cite{Lu:Hamilton}; for a recent treatment, see
MacKinnon \& Macpherson \cite{MacKinnon:Macpherson} and references therein
) and to transport phenomena in magnetically confined plasmas (Newman et
al. \cite{Newman:Carreras},
Carreras et al. \cite{Carreras:Newman},
Dendy \& Helander \cite{Dendy:Helander}, \cite{Dendy:Hel2}). It is clearly of interest to
establish whether,
in certain circumstances, the detailed modelling of astrophysical accretion
flows
could be substituted by a simple SOC sandpile paradigm.  This is a highly
cross-disciplinary question.  In the present paper, we aim to carry forward the
debate in two ways: by identifying in greater detail the classes of accretion
flow where SOC might play a role; and by examining theoretical arguments
that we
believe point clearly towards SOC in some accretion flows.

\section{Some sites of interest}

It was noted by Bak et al.  (\cite{Bak:Tang}) that flickering signals with $1/f$ 
power spectra have been observed for the X-ray variability of active galactic 
nuclei (AGNs), specifically: 0.05-2keV X-rays from the Seyfert galaxy NGC4051 
(Lawrence et al \cite{Lawrence:Watson}) and 2-7keV X-rays from the Seyfert 
galaxy NGC5506 (McHardy \& Czerny \cite{McHardy:Czerny}), and for 10-140 keV 
X-rays from the massive compact binary Cyg~X-1 (Nolan et al.  
\cite{Nolan:Gruber}).  Similar $1/f$ spectra for X-ray variability from binary 
accreting systems were noted by Mineshige et al.  (\cite{Mineshige:Takeuchi}) 
(neutron stars, Makashima \cite{Makashima:1988}) and by Geertsema \& Achterberg 
(\cite{Geertsema:Achterberg}) (cataclysmic variables and dwarf novae, Wade \& 
Ward \cite{Wade:Ward}).  While we believe this list can be extended, as we 
discuss below, it is also important to be more specific about where, in these 
diverse systems, SOC might be occurring.

Perhaps the simplest case is presented by a paradigmatic AGN. Let us take for 
this the standard picture of a massive black hole fed by a cascade of 
structures: the accretion disc, the molecular torus and on a larger scale the 
galactic disc and its barred or spiral structure, which together generate a 
clumpy and irregular transfer of gas and stars.  This irregular mass transfer 
could be analogous to the random and discrete feeding of a sandpile with grains 
of sand near its apex.  It is believed that, at least in thin discs, the 
residence time $\tau_{res}$ of matter in the disc is much longer than 
the free--fall time, $\tau_{ff}\approx (R^3/GM)^{1/2}$ at radius $R$:
\[\frac {\tau_{res}} {\tau_{ff}}\sim \frac {R^2} {\alpha H^2}\]
where $H\ll R$ is the thickness of the disc, and $\alpha\ll 1$ is the 
well-known viscosity parameter. The condition $\tau_{res}\gg\tau_{ff}$, 
which is implicit in most accretion disc models, appears to be 
necessary, at least in principle, for the validity of a sandpile--type 
approach to mass transfer within the disc.

Thus if the accretion disc, like the sandpile, is in a state 
of SOC, there would be no clear link between the pattern of accretion from the 
torus and the pattern of avalanches leading to mass transfer across the disc, 
over its inner edge, and onto the black hole, giving rise to the X-ray signal.  
The latter would automatically display $1/f$ flicker.  On larger space and 
timescales relating to our own Galaxy, even steady gas inflow from the Galactic 
bar could also result in avalanches to the inner regions, and these avalanches 
could give rise to episodes of AGN activity during an otherwise quiescent phase, 
with a duty cycle of a few percent, as inferred on statistical grounds for 
other AGNs (Mezger et al., \cite{Mezger:Duschl}).  The sandpile-SOC paradigm thus provides a 
candidate framework for summing up the consequences of the complex and intricate 
physics that relates the largescale dynamics of the Galactic disc to the 
activity of its central black hole.

Cataclysmic variables, and in particular dwarf novae, present further 
opportunities for SOC. First, the flow of matter from the secondary 
across the inner Lagrangian point could itself be an avalanching process 
governed by SOC. In this case, if the radiation source was a hot spot 
where the accreting mass flow reached the outer edge of an accretion 
disc, it would flicker with $1/f$ statistics.  Steady mass flow from the 
secondary would also be compatible with an SOC signal, however, provided 
the latter originated from SOC mass transfers from the accretion disc 
(or accretion column in the strongly magnetised regime) to the white 
dwarf.  Either regime appears possible in principle, both for dwarf 
novae and in the wider context of binary accreting systems.  In dwarf 
novae, it is pointed out by van Amerongen et al.  
(\cite{Amerongen:Kuulkers}) and Lasota et al.  (\cite{Lasota:Hameury}) 
that outbursts are due to suddenly increased accretion towards the white 
dwarf, but that it is unclear whether the instability resides in mass 
transfer from the secondary to the accretion disc, or across the 
accretion disc and onto the white dwarf.  We note that, observationally, 
flickering is more often strong during the quiescent phase of dwarf 
novae than during major eruptions (Wade \& Ward \cite{Wade:Ward}).  
Questions relating to the location of unstable flows in soft X-ray 
transients (Lasota et al.  \cite{Lasota:Narayan}) are similar to those 
in dwarf novae.  For the wind-fed X-ray binary pulsar GX301-2, a model 
has been proposed (Orlandini \& Morfill \cite{Orlandini:Morfill}) 
involving ``noisy" accretion of blobs of matter formed by 
magnetohydrodynamical (MHD) instability at the magnetospheric radius, 
and not caused by inhomogeneities present in the stellar wind from the 
optical companion.  This approach is somewhat reminiscent of that of 
Baan (\cite{Baan:1977}), where accreted matter accumulates at the 
magnetopause of a rotating neutron star until an interchange instability 
is triggered, after which the released matter generates an X-ray burst.  
Baan (\cite{Baan:1977}) suggested that, since most of the time between 
bursts is a refilling time, an approximately linear relation should 
exist between burst energy and the subsequent quiescent interval.  
However, in an SOC model where randomly arising local instability is 
sufficient to trigger a global avalanche, there would be no such 
correlation.  This appears to be a key observational discriminant for 
the possible presence of SOC in a given accreting system.  We also note 
that, from the theoretical point of view, Frank {\it et al.} 
\cite{Frank:King} have pointed out that a local instability in a given 
annulus of the disc can only trigger large--scale instability accross 
the disc if parameters in neighbouring annuli are such that the effect 
of this instability in these annuli can in turn trigger local 
instability there. This amounts to a prescription for a sandpile--type 
approach, and hence for the possibility of SOC.

It seems clear from the foregoing that there is good observational and
interpretative motivation for testing for SOC in a broad range of accreting
astrophysical systems, encompassing flickering AGNs and certain distinct
locations within a variety of binary objects.  Firm identification of SOC would
yield information on the global consequences of the smaller-scale physics of
the accretion process, while short-circuiting the need for detailed modelling.
Let us now turn to theoretical arguments for expecting SOC.

\section{Theoretical considerations}

The identification of mechanisms responsible for the dissipation of shear flow
energy within an accretion disc is an active field of research, because of its
importance in determining transport across the disc and onto the compact
object.
If, as has been suggested (Bak et al \cite{Bak:Tang}, Mineshige et al.
\cite{Mineshige:Takeuchi}), the combined global effects of local transport
physics result in SOC, observational signatures of the type described above
will
emerge.  It is widely accepted (see, for example, Longair (\cite{Longair:1994})
and Narayan (\cite{Narayan:1997})) that anomalous viscosity caused by MHD
turbulence probably plays an important role in the flux of angular momentum
within accretion discs. MHD turbulence can arise naturally in accretion
discs, see for example the
instability mechanisms proposed by Tagger et al. (\cite{Tagger:Henriksen}),
Vishniac et al. (\cite{Vishniac:Jin}), and Balbus \& Hawley
(\cite{Balbus:Hawley}).  It has also been pointed out (Chen et al.
\cite{Chen:Abramowicz}) that there exist ranges of accretion rate $\dot{M}$ and
disc radius $R$ for which no stable steady state solution of the basic
equilibrium equations is possible.  In this case, it is suggested (Chen et al.
\cite{Chen:Abramowicz}, Narayan \cite{Narayan:1997}) that the flow is
forced into
a time-dependent variable mode, satisfying the required $\dot{M}$ in the mean.
This provides further motivation to test the applicability of sandpile-type
models.  In this section, we concentrate on the model for MHD turbulent energy
dissipation in accretion discs presented by Geertsema \& Achterberg
(\cite{Geertsema:Achterberg}), because it appears to
give rise to SOC.  While the link to cataclysmic variable and dwarf nova
observations (but not, explicitly, SOC) was made by  Geertsema \& Achterberg
(\cite{Geertsema:Achterberg}) themselves, we believe that this work has much
wider implications: for the general question of SOC in accretion flows; for the
connection between SOC and turbulence in general;
for the role of SOC in plasma physics, since it represents the first
instance where SOC
has been observed in a mathematical model derivable from the fundamental
equations of MHD; and for the role of SOC in terrestial experimental systems -
real sandpiles and ricepiles, as distinct from mathematical idealisations
thereof
- where uncertainties about its scope remain (Nagel \cite{Nagel:1992}, Feder
\cite{Feder:1995}, Frette et al. \cite{Frette:Christensen}, Christensen et al.
\cite{Christensen:Corral}).

Before considering the model of Geertsema \& Achterberg 
(\cite{Geertsema:Achterberg}) in greater detail, let us turn to its 
results.  Figure 12 of Geertsema \& Achterberg 
(\cite{Geertsema:Achterberg}) shows the calculated times series of 
energy dissipation events within the disc.  We note that this is 
qualitatively very similar to the observed time series of energy 
dissipation measured in an experimental ricepile displaying SOC, Fig 2c 
of Frette et al.  (\cite{Frette:Christensen}), and in a related 
mathematical model of Dendy \& Helander (\cite{Dendy:Helander}, fig.  3, 
and \cite{Dendy:Hel2}, fig.5).  More quantitatively, Fig 13 of Geertsema 
\& Achterberg (\cite{Geertsema:Achterberg}) shows the power spectrum of 
energy dissipated by MHD disc turbulence, which displays the $1/f$ 
dependence characteristic of SOC; compare, for example, Fig 3 of Frette 
et al.  (\cite{Frette:Christensen}) and Fig 3 of Christensen et al.  
(\cite{Christensen:Corral}) which show measured spectra of energy 
dissipation and particle transit times, respectively, in SOC ricepiles.

Given the clear indications of SOC emerging from the MHD turbulence model of
Geertsema \& Achterberg (\cite{Geertsema:Achterberg}), it will be of interest
to  establish  how  it  has arisen.  A full explanation must await diagnostic
analysis of the code runs generated by this model.  Pending this, we
conclude the
present section by seeking to identify some of the relevant salient features.
In outline, the model of Geertsema \& Achterberg (\cite{Geertsema:Achterberg})
is constructed  as  follows.

The accretion disc is regarded as a differentially rotating turbulent MHD
fluid,
and is modelled by a reduced system of equations reflecting the most important
features of three-dimensional MHD turbulence. The disc is assumed to be thin in
comparison with its diameter, and the flow is taken to be subsonic and hence
incompressible. The possible existence of a large-scale magnetic
field is neglected, and the only turbulent structures considered have a length
scale shorter than the height of the disc, making the turbulence essentially
three-dimensional. In a coordinate system rotating with the disc angular
velocity
${\vek \Omega}= {\bfm e}_z \Omega_K(R)$ at radius $r=R$, the simplified MHD
equations for the flow velocity ${\vek u}$ and the normalised magnetic field
vector ${\vek b} = {\vek B} / (4 \pi \rho)^{1/2}$ are
\begin{eqnarray}
\dot{\vek u} = -{\vek \nabla}\Pi - 2 {\vek \Omega} &\times& {\vek u} -
\left( r \frac{d\Omega_K}{dt} \right)_R u_x {\vek e}_y \nonumber \\
&+& \nu \nabla^2 	{\vek u} + {\vek b} \cdot \nabla {\vek b}
- {\vek u} \cdot \nabla {\vek u}, \nonumber
\end{eqnarray}
\[
\dot{\vek b} = \left( r \frac{d\Omega_K}{dt} \right)_R b_x {\vek e}_y
	- \eta \nabla^2 {\vek b} + {\vek b} \cdot \nabla {\vek u}
	- {\vek u} \cdot \nabla {\vek b},
\]
where $x=r-R$ and $y=R\phi$ are the local radial and azimuthal
coordinates respectively, $\eta$ is the resistivity, and $\nu$ is the
viscosity coefficient. $\Pi = P/\rho + b^2/2$ is the total scalar 
pressure, while the ${\vek b} \cdot \nabla {\vek b}$ term contains the 
off--diagonal terms of the MHD stress tensor.

As these equations are still too complex for a numerical treatment over a
sufficiently large dynamic range in scales, they were simplified further
following a suggestion by Desnyanski \& Novikov (\cite{Desnyanski:Novikov}) in
the context of hydrodynamics, and later applied to MHD by Gloaguen et al.
(\cite{Gloaguen:Leorat}). In this approximation, the
space of wave vectors $\vek k$ is discretized into a finite set of $k_n$,
and the non-linear
interaction between different components of the Fourier transforms of
${\vek u}$ and ${\vek b}$ is described by the set of equations
\begin{eqnarray}
\frac{du_n}{dt} + \nu k_n^2 u_n &=&
	\alpha k_n \left( u_{n-1}^2 - 2 u_n u_{n+1} \right) \nonumber \\
&+&	\beta k_n  \left( u_{n-1} u_n - 2 u_{n+1}^2 \right) - ( u
\leftrightarrow b ),
\nonumber
\end{eqnarray}
\begin{eqnarray}
\frac{db_n}{dt} + \eta k_n^2 b_n &=&
\alpha k_n \left( u_{n-1} b_{n-1} - 2 u_n b_{n+1} \right) \nonumber \\
&+&
\beta k_n  \left( u_{n-1} b_n - 2 u_{n+1} b_{n+1} \right) -
( b \leftrightarrow u ). \nonumber
\end{eqnarray}
In the three-dimensional generalisation of this approximation scheme,
the discretisation is made
by dividing the ${\vek k}$-space into into spherical shells, thus
discarding the information  regarding the direction of
${\vek k}$. This allows a greatly simplified system of nonlinear equations
to be written down, analogous to that of Gloaguen et al.
(\cite{Gloaguen:Leorat}), which are supplemented with additional terms to
account for the effects of differential rotation.

These equations, which are taken to model the turbulent cascade of MHD, 
were solved numerically by Geertsema \& Achterberg 
(\cite{Geertsema:Achterberg}).  They found that the turbulent shear 
stress can be very large, and has large, chaotic fluctuations on time 
scales of a few rotation periods.  Perhaps the most striking feature of 
the simulations is, however, that the dissipation of energy at the 
smallest scales of the turbulent cascade is very intermittent.  Energy 
is released in avalanches with a wide range of sizes, and the power 
spectrum of the dissipation rate obeys a $1/f$ power law over nearly two 
orders of magnitude in intensity, see their Fig 13.  This behaviour is, 
as already stated, similar to that of simple mathematical sandpile 
models.  While a power law is to be expected from any scale-free model, 
in particular a Kolmogorov-type one in the inertial range, we note that 
the similarities between this system and mathematical sandpiles 
apparently extend further.  Both are fundamentally governed by 
nearest-neighbour interactions between a discrete number of nodes.  In 
the MHD accretion model, these reside in ${\vek k}$-space, so the 
interaction is between wave modes with similar wavelengths rather than 
between adjacent regions in real space.  The dissipation is provided by 
viscosity and resistivity at large $k$, resembling the removal of 
material from the edge of a sandpile.  The MHD accretion model is, of 
course, much more complex than the simple sandpile algorithms considered 
so far in the literature.  Two fields, $\vek u$ and $\vek b$, each with 
three components, are involved rather than the single height parameter 
of conventional sandpile algorithms, and the time evolution is governed 
by differential equations rather than difference equations.  
Nevertheless, both the MHD accretion disc and simple sandpile models 
appear to exhibit similar self-organised, critical behaviour, supporting 
the claim often made in the sandpile literature that SOC is universal 
phenomenon shared by large classes of cellular automata.

\section{Discussion}

We have shown in this paper that the numerical simulations by Geertsema \&
Achterberg (\cite{Geertsema:Achterberg}) of viscous resistive MHD turbulence in
an accretion disc give rise to behaviour characteristic of self-organised
criticality in a sandpile.  This similarity may help to explain certain
observed properties of a range of accreting astrophysical systems, which we
have reviewed.  Furthermore, the result is of intrinsic scientific interest as
an ab initio demonstration of the emergence of SOC from a system of MHD-based
equations.  Further analysis of the numerical results, to establish how the
SOC-sandpile phenomenology arises, would be of great interest both for
accretion disc astrophysics and for fundamental plasma physics.  At the present
stage of development of realistic MHD simulations (for example of the link
between smallscale dissipation and the time behaviour of the flux of matter
onto the central object), direct numerical proof of the occurence of
SOC-sandpile phenomenology is still a remote objective.  However, the capacity
of the SOC-sandpile paradigm to circumvent complex ``full mathematical
descriptions" is the basis for its present attraction in many fields of physics.
If future work confirms the importance of SOC-sandpile phenomenology in
accreting systems that is suggested in the present paper, it would permit a
dissociation (at least at zeroth order) of the detailed physics of turbulence
in the disc from the global modelling of this class of astrophysical object.

\begin{acknowledgements}
This work was supported in part by the Commission of the European Communities
under Contracts ERBCHRXCT940604 and FMRX-CT98-0168. The authors are 
grateful to Bram Achterberg for helpful comments as a referee.
 \end{acknowledgements}

\end{document}